\documentclass[a4paper,twocolumn,english,amsmath,amssymb,nofootinbib,showpacs,preprintnumbers]{revtex4}

\makeatletter

\usepackage[T1]{fontenc}
\usepackage[latin1]{inputenc}
\usepackage{graphicx}
\usepackage{amssymb}
\usepackage{ae, aecompl}
\usepackage{dcolumn}% Align table columns on decimal point
\usepackage{bm}% bold math
\usepackage{graphics}
\usepackage{epsfig}
\usepackage{babel}

\makeatother

\begin{document}

\preprint{TUM-HEP-633/06}

\title{Quantum Corrections in Quintessence Models}

\author{Mathias Garny}

\email{mgarny@ph.tum.de}

\affiliation{Physik-Department T30d, Technische Universit\"{a}t M\"{u}nchen\\
 James-Franck-Stra\ss{}e, 85748 Garching, Germany}

\begin{abstract}
We investigate the impact of quantum fluctuations on a light rolling quintessence field
from three different sources, namely, from a coupling to the
standard model and dark matter, from its self-couplings and from its coupling to gravity.
We derive bounds for time-varying masses from the 
change of vacuum energy,
finding $\Delta m_e/m_e\ll 10^{-11}$ for the electron
and $\Delta m_p/m_p\ll 10^{-15}$ for the proton since redshift $z\sim 2$, whereas the neutrino
masses could change of order one. 
Mass-varying dark matter is also constrained.
Next, the self-interactions are investigated.
For inverse power law potentials, the effective potential
does {\it not} become infinitely large at small field values, but saturates at a finite maximal value.
We discuss implications for cosmology.
Finally, we show that one-loop corrections  induce
non-minimal gravitational couplings involving arbitrarily 
high powers of the curvature scalar $R$,
indicating that quintessence entails modified gravity effects.
\end{abstract}

\pacs{95.36.+x, 04.62.+v, 11.10.Gh}

%\vspace*{0cm}

\maketitle

%%%%%%%%%%%%%%%%%%%%%%%%%%%%%%%%%%%%%%%%%%%%%%%%%%%%%%%%%%%%%%%%%%%%%%%%%%%%%%%%%%%%%%%%%%%%%%%%%%%%%%%%%%%%%%%%%%%%%%%%%%%%%%%%%%%%%%

\section{Introduction}

%%%%%%%%%%%%%%%%%%%%%%%%%%%%%%%%%%%%%%%%%%%%%%%%%%%%%%%%%%%%%%%%%%%%%%%%%%%%%%%%%%%%%%%%%%%%%%%%%%%%%%%%%%%%%%%%%%%%%%%%%%%%%%%%%%%%%%

A possible explanation for the observed \cite{Riess:1998cb,Perlmutter:1997zf} acceleration
of the universe is given by a light rolling scalar field \cite{Wetterich:1987fm,Ratra:1987rm},
usually called quintessence field. The dynamics of this field can lead to a decaying dark energy,
and thus address the question why the cosmological ``constant" is small but non-zero today.
Presently, there is a huge number of dark energy models based on  scalar fields,
see e.g. ref. \cite{Copeland:2006wr} for a review.
Models admitting  e.g. tracking solutions \cite{Steinhardt:1999nw} 
or general scaling solutions \cite{Copeland:2006wr} additionally possess some
appealing properties like attractors which wipe out the dependence on the initial conditions
of the field in the early universe, or a dynamical mechanism naturally yielding an extremely small classical mass
of the quintessence field of the order of the Hubble parameter.
The latter can be necessary e.g. to inhibit the growth of inhomogeneities of the scalar field \cite{Ratra:1987rm}. 

However, the rolling quintessence field usually cannot be regarded as completely independent
of other degrees of freedom. If the quintessence dynamics
are for example governed by a low-energy effective theory which is determined by
integrating out some unknown high energy degrees of freedom,
involving e.g. quantum gravity, string theory or supergravity \cite{Copeland:2006wr,Chung:2002xj}, the low-energy
theory should generically contain couplings and self-couplings of the quintessence field suppressed by
some large scale, e.g. the Planck scale. In some cases such couplings can be directly constrained
observationally, like for a coupling to standard model gauge fields \cite{Carroll:1998zi},
whereas a significant interaction with dark matter seems to be possible \cite{Amendola:2002bs} and is used in many models, e.g. 
\cite{Amendola:2000uh,Farrar:2003uw,Zhang:2005rg,Huey:2004qv,Srivastava:2004xt,Zimdahl:2001ar},
often accompanied by a varying dark matter mass (VAMP) 
\cite{Rosenfeld:2005pw,Franca:2003zg,Comelli:2003cv,Hoffman:2003ru}.
Models leading to time-varying standard model masses and couplings, including mass-varying neutrinos (MaVaNs), 
from a corresponding coupling to a rolling dark energy field are also frequently considered, see e.g. 
\cite{Wetterich:2002wm,Doran:2002ec,Doran:2002bc,Fardon:2003eh,Chiba:2001er,Wetterich:2003jt,Anchordoqui:2003ij,Lee:2004vm,Copeland:2003cv,Brax:2006dc},
which can lead to potentially observable effects like a variation of the electron to proton
mass ratio \cite{Reinhold:2006zn,Ivanchik:2005ws} and the fine-structure constant \cite{Webb:2000mn,Chand:2004ct} 
or violations of the equivalence principle \cite{Wetterich:2003jt,Uzan:2002vq},
and could have an effect on BBN \cite{Campbell:1994bf,Uzan:2002vq}.
Furthermore, higher order self-interactions of the field seem to be a typical feature
necessary for a successful dark energy model, involving e.g. exponentials or inverse powers
of the field \cite{Ratra:1987rm,Wetterich:1987fm,Albrecht:1999rm,Amendola:1999er,Steinhardt:1999nw,Binetruy:1998rz}. 
Non-minimal gravitational couplings of the scalar field have also been studied in various
settings \cite{Perrotta:1999am,Chiba:1999wt,Carvalho:2004ty,deRitis:1999zn,Faraoni:2000wk,Catena:2004ba,Boisseau:2000pr,Gannouji:2006jm},
constrained e.g. by solar system tests of gravity and BBN.

Because of the presence of quantum fluctuations of the standard model degrees of freedom
as well as of dark matter and of dark energy itself, the dynamics of the quintessence field
receive radiative corrections. Therefore, it is important to study the
robustness against these corrections, see e.g. refs. 
\cite{Kolda:1998wq,Brax:1999yv,Doran:2002bc,Uzan:1999ch,Bartolo:1999sq,Riazuelo:2001mg,Onemli:2002hr} for previous work.
Apart from the long-standing problem of the overall normalization of the effective
quintessence potential (i.~e. the ``cosmological
constant problem"), which is not addressed here,
quantum corrections can influence the dynamics e.g. by distorting the {\it shape} or
the flatness of the scalar potential.
In the case of a coupling
to standard model or heavy cold dark matter particles, this leads to tight upper bounds
for the corresponding couplings,
due to a physically relevant field-dependent  shift 
in the corresponding contribution to the vacuum energy \cite{Donoghue:2001cs,Banks:2001qc}.
The quantitative bounds obtained in this way can be translated into bounds for time-varying masses 
and for the coupling strength to a long-range fifth force mediated by the quintessence field,
as will be shown in section \ref{sec:2}.
The self-coupling and gravitational coupling
of the dark energy field are necessary ingredients of basically any given model. The corresponding quantum corrections 
can lead to significant modifications of the shape of the scalar potential, discussed in section \ref{sec:3}
using an infinite resummation of bubble diagrams. This accounts for the fact that the effective
theory for a scalar field does not decouple from the high energy regime
due to the presence of quadratic divergences. We also discuss implications for cosmology.
In section \ref{sec:4} we investigate which kind of non-minimal gravitational couplings 
are induced by quantum fluctuations of the dark energy scalar field.

%%%%%%%%%%%%%%%%%%%%%%%%%%%%%%%%%%%%%%%%%%%%%%%%%%%%%%%%%%%%%%%%%%%%%%%%%%%%%%%%%%%%%%%%%%%%%%%%%%%%%%%%%%%%%%%%%%%%%%%%%%%%%%%%%%%%%%

\section{\label{sec:2}Lifting of the scalar Potential}

%%%%%%%%%%%%%%%%%%%%%%%%%%%%%%%%%%%%%%%%%%%%%%%%%%%%%%%%%%%%%%%%%%%%%%%%%%%%%%%%%%%%%%%%%%%%%%%%%%%%%%%%%%%%%%%%%%%%%%%%%%%%%%%%%%%%%%

Generically, the light mass of the quintessence field
is unprotected against huge corrections induced by the
quantum fluctuations of heavier degrees of freedom.
Furthermore, not only the mass but also the total potential
energy have to be kept small, which is the ``old cosmological constant problem''.
In General, the explanation of the present acceleration and the question about 
huge quantum field theoretic contributions to the cosmological constant
may have independent solutions. However, it is required in quintessence models
that the total cosmological constant is small.

Here we will take the following attitude: Even if we accept
a huge amount of fine-tuning and choose the quintessence
potential energy and mass to have the required values {\it today}
by a suitable renormalization, there may be huge corrections
to the potential at a value of the quintessence field
which is slightly displaced from todays value. Since the
scalar field is rolling, such corrections would affect
the behaviour of the quintessence field in the past,
and could destroy some of the desired features  (like tracking behaviour)
of dynamical dark energy.  

To calculate the effect of quantum fluctuations,
we will impose suitable renormalization conditions for the
effective quintessence potential. Therefore, we are going to argue
that under certain general prerequisitions, there remain only three
free parameters (linked to the quartic, quadratic and logarithmic divergences)
that can be used to fix (or fine-tune) the effective potential
which is induced by the fluctuations of heavier particles coupled to the quintessence field.
Following the above argumentation, these free parameters
will be fixed at one-loop level by imposing the renormalization condition that the quantum
contributions to the effective potential and its first and second
derivative vanish today ($\phi_0\equiv\phi(t_0)$):
\begin{eqnarray}\label{RenCond}
V(\phi=\phi_0)_{1L} & = & 0 \nonumber\\
V'(\phi=\phi_0)_{1L} & = & 0 \\
V''(\phi=\phi_0)_{1L} & = & 0, \nonumber
\end{eqnarray}
where $V(\phi)_{1L}$ denotes the one-loop contribution to the effective potential
$V_{\textit{eff}}(\phi)\equiv V(\phi)+V_{1L}(\phi)+\dots$. 

Since the quintessence field generically changes only slowly on cosmological time-scales,
one expects that the leading effect of
quantum fluctuations is suppressed by a factor of the order 
\begin{equation}
V'''(\phi=\phi_0)_{1L}(\dot\phi(t_0)\,\Delta t)^3
\end{equation}
(with $\Delta t$ of the order of a Hubble time) compared to the potential $V(\phi)$.

The coupling between quintessence and any massive particle species $j$
is modeled by assuming a general dependence of the mass on the quintessence field.
This general form includes many interesting and potentially observable possibilities,
like a time-varying (electron- or proton-) mass $m_j(\phi(t))$,
a Yukawa coupling $dm_j/d\phi$ to
fermions (e.g. protons and neutrons) mediating a new long-range fifth force, 
or a coupling between dark energy and dark matter ({\textit{dm}})
of the form (see e.g. \cite{Amendola:2002bs})
\begin{equation}
\dot\rho_{dm} + 3H\rho_{dm} = \rho_{dm}\frac{d\,\ln m_{dm}(\phi)}{d\phi} \dot\phi.
\end{equation}
In terms of particle physics, a dependence of the mass on the
dark energy field $\phi$ could be produced in many ways,
which we just want to mention here.
One possibility would be a direct $\phi$-dependence of the Higgs Yukawa couplings or of 
the Higgs VEV. For Majorana neutrinos, 
the Majorana mass of the right-handed neutrinos could depend on $\phi$ leading to varying
neutrino masses via the seesaw mechanism \cite{Gu:2003er}. The mass of the proton and neutron
could also vary through a variation of the QCD scale, for example induced by
a $\phi$-dependence of the GUT scale \cite{Wetterich:2002ic}. Additionally, a variation of the
weak and electromagnetic gauge couplings could directly lead to a variation of the
radiative corrections to the masses \cite{Donoghue:2001cs}. Possible parameterizations
of the $\phi$-dependence are $m(\phi)=m_0(1+\beta f(\phi/M_{pl}))$ with a dimensionless
coupling parameter $\beta$ and a function $f(x)$ of order unity or $m(\phi)=m_0\exp(\beta\phi/M_{pl})$ \cite{Doran:2002bc}.

\subsection*{Induced Effective Potential}

The one-loop contribution to the effective potential for the quintessence field can be calculated 
in the standard way from the
functional determinants of the propagators with mass $m(\phi)$:
\begin{equation}\label{VacBubble}
\begin{array}{ll}
\displaystyle V_{1L}(\phi) = \frac{1}{2}\int\!\!\!\frac{d^4q}{(2\pi)^4} & \!\!\!\displaystyle\left(\sum_B g_B \ln(q^2+m_B(\phi)^2) \right. \\ 
&\displaystyle - \left. \sum_F g_F \ln(q^2+m_F(\phi)^2)\right), 
\end{array}
\end{equation}
where $B$ and $F$ run over all bosons and fermions with internal degrees of freedom $g_B$ and $g_F$ respectively,
and the momentum has been Wick-rotated to euclidean space.
To implement the renormalization conditions (\ref{RenCond}), we consider the class of integrals
\begin{equation}
\begin{array}{l}\label{Ik}
\displaystyle I_0(m^2)  \displaystyle\equiv  \displaystyle\int\!\!\!\frac{d^4q}{(2\pi)^4} \ln (q^2+m^2) \\
\displaystyle I_k(m^2)  \displaystyle\equiv  \displaystyle\int\!\!\!\frac{d^4q}{(2\pi)^4} \frac{1}{(q^2+m^2)^k} = \frac{(-1)^{k-1}}{(k-1)!}\frac{d^k}{(dm^2)^k} I_0(m^2),
\end{array}
\end{equation}
which are finite for $k\geq 3$. 
Following the standard procedure described e.g. in \cite{Weinberg:1995mt} the divergences in $I_0$, $I_1$ and $I_2$
are isolated by integrating $I_3$ with respect to $m^2$, yielding
\begin{equation}\label{I0}
\begin{array}{l}
\displaystyle I_0(m^2) = 2 \int^{m^2}\!\!\!dm_3^2\int^{m_3^2}\!\!\!dm_2^2\int^{m_2^2}\!\!\!dm_1^2 \ I_3(m_1^2) \\
\displaystyle \hspace*{2cm}\rule{0mm}{4mm} + D_0 + D_1 m^2 + D_2 m^4,
\end{array}
\end{equation}
with infinite integration constants $D_0$, $D_1$ and $D_2$. Thus one is led to introduce
three counterterms proportional to $m^0$, $m^2$ and $m^4$ to cancel the divergences,
which can be easily reabsorbed by a shift of the scalar potential $V$.
This leaves a finite part $I_0^{\rm finite}$ of the same form as (\ref{I0})
but with the three infinite constants replaced by three finite parameters that have to be fixed
by the three renormalization conditions (\ref{RenCond}). It is easy to see that the appropriate
choice can be expressed by choosing the lower limits in the integration over the mass $m^2$
to be equal to its todays value $m_0^2$:
\begin{equation}\label{I0finite}
\begin{array}{l}
\displaystyle I_0^{\rm finite}(m^2;m_0^2) = 2 \int^{m^2}_{m_0^2}\!\!\!dm_3^2\int^{m_3^2}_{m_0^2}\!\!\!dm_2^2\int^{m_2^2}_{m_0^2}\!\!\!dm_1^2 \ I_3(m_1^2) \\
\displaystyle \qquad \rule{0mm}{7mm}= \frac{1}{32\pi^2}\left(m^4\left(\ln\frac{m^2}{m_0^2}-\frac{3}{2}\right)+2m^2m_0^2-\frac{1}{4}m_0^4\right),
\end{array}
\end{equation}
where $I_3(m^2)=1/(32\pi^2m^2)$ has been used.

Thus the renormalized one-loop contribution to the effective potential which fulfills the renormalization conditions (\ref{RenCond})
is uniquely determined to be
\begin{equation}
\begin{array}{lcl}\label{V1L}
\displaystyle V_{1L}(\phi) & \displaystyle= & \!\!\!\displaystyle\frac{1}{2}\left(\sum_B g_B I_0^{\rm finite}(m_B(\phi)^2;m_B(\phi_0)^2) \right. \\
&& \displaystyle \left. - \sum_F g_F I_0^{\rm finite}(m_F(\phi)^2;m_F(\phi_0)^2)\right). 
\end{array}
\end{equation}
The effective potential renormalized in this way can be regarded
as the result of a fine-tuning of the contributions from the quantum fluctuations
of heavy degrees of freedom to the quintessence potential energy, slope and mass
at its todays values, i.e. evaluated for $\phi=\phi_0$. However,
when the quintessence VEV had different values in the cosmic history,
the cancellation does not occur any more and one expects the huge corrections
of order $m^4$ to show up again, unless the coupling is extremely weak.
Indeed, this argument yields extremely strong bounds for the variation
of the masses with the rolling field $\phi$. Similar considerations
have been done e.g. in \cite{Donoghue:2001cs,Banks:2001qc}. To
obtain a quantitative limit we require that the one-loop contribution to the potential
should be subdominant during the relevant phases of cosmic history up to now,
which we take to be of the order of a Hubble time, in order to ensure that the
quintessence dynamics, e.g. tracking behaviour, are not affected. For the corresponding
$\phi$-values this means that we require
\begin{equation}\label{BoundForV1L}
V_{1L}(\phi)  \ll V(\phi).
\end{equation}
If we Taylor-expand the one-loop effective potential (\ref{V1L}) around todays value $\phi_0$,
the first non-vanishing contribution is by construction of third order,
\begin{equation}
\begin{array}{l}\label{V1LTaylor}
\displaystyle V_{1L}(\phi) \displaystyle\approx \displaystyle\frac{1}{3!}V_{1L}'''(\phi_0)(\phi-\phi_0)^3 \\
\quad \displaystyle\approx  \displaystyle\frac{1}{3!}\frac{1}{32\pi^2}\sum_j\frac{\pm g_j}{m_j(\phi_0)^2}\left(\frac{dm_j^2}{d\phi}(\phi-\phi_0)\right)^3 \\
\quad \displaystyle\approx  \displaystyle\frac{1}{96\pi^2}\sum_j \pm g_j m_j(\phi_0)^4 \left(\frac{d\ln m_j^2}{d\ln V''}\ln\frac{V''(\phi)}{V''(\phi_0)}\right)^3.
\end{array}
\end{equation}
Here the index $j$ runs over bosons $B$ and Fermions $F$ (with the minus sign in front of $g_j$ for the latter), and eq. (\ref{I0finite})
has been used. In the last line, we have rewritten the dependence on the quintessence field $\phi$
in a dependence on its mass $m_\phi^2(\phi)\equiv V''(\phi)$. Typically, the mass is of the order of the Hubble parameter,
which is today $H_0\sim 10^{-33}eV$. In many generic scenarios, e.g. for tracking quintessence models \cite{Steinhardt:1999nw},
the quintessence mass also scales proportional to the Hubble parameter $H$ during cosmic evolution.
Therefore, we assume that 
\begin{equation}\label{MassRedshiftRel}
\ln V''(\phi)/V''(\phi_0) \sim \ln H^2/H_0^2 \lesssim 3\ln(1+z),
\end{equation}
where $z$ is the cosmic redshift. If we want the inequality (\ref{BoundForV1L}) to hold 
up to a redshift $z_{\textit{max}}$,
the most conservative assumption is to replace the logarithm in the last line in (\ref{V1LTaylor}) by its maximal value
of order $3\ln(1+z_{\textit{max}})$ and the right hand side of (\ref{BoundForV1L}) by the minimal value $V(\phi_0)$.
Furthermore, the  inequality (\ref{BoundForV1L}) is certainly fulfilled if each individual contribution
to the one-loop potential (\ref{V1L}) respects it. 
Altogether, under these assumptions the requirement (\ref{BoundForV1L}) 
that the quintessence dynamics are unaltered up to a redshift $z_{\textit{max}}$ yields the bound
for the variation of the mass $m_j$ of a species $j$ (with $g_j$ internal degrees of freedom)
with the quintessence mass scale $V''\sim H^2$
\begin{equation}\label{BoundOnMassVar}
 \left|\frac{d\ln m_j^2}{d\ln V''}\right| \ll \frac{1}{3\ln(1+z_{\textit{max}})}\left(\frac{96\pi^2 V(\phi_0)}{g_j m_j(\phi_0)^4}\right)^\frac{1}{3}.
\end{equation}
This bound is the main result of this section. It scales with mass like $m^{-4/3}$,
i.e. the bound gets tighter for heavier particles. Inserting $z_{\textit{max}}\sim z_{eq}\sim 10^3$ and
expressing the potential energy 
$
V(\phi_0)=\frac{1-\omega_{de}}{2}\Omega_{de}\frac{3H_0^2}{8\pi G}
$
in terms of the dark energy fraction $\Omega_{de}$ and equation of state $\omega_{de}$
with $H_0\sim 70{\rm km}/{\rm s\, Mpc}$
yields
\begin{equation}\label{BoundOnMassVar2}
\left|\frac{d\ln m_j^2}{d\ln V''}\right| \ll \left(\frac{1-\omega_{de}}{2}\frac{\Omega_{de}}{0.7}\right)^{\frac{1}{3}} \frac{1}{\sqrt[3]{g_j}}
\left(\frac{1.3{\rm meV}}{m_j(\phi_0)}\right)^{\frac{4}{3}}.
\end{equation}
Finally, we want to remark that there remains the possibility
that several masses $m_j$ change in such a way that the {\it total}
contribution to the effective potential stays small \cite{Donoghue:2001cs}.
It would be interesting to look for a special (dynamical) mechanism or a symmetry which leads to such
fine-tuned correlated changes. Otherwise, there seems to be no motivation for such a behaviour.
An example for such a mechanism could be based on supersymmetry, where the masses
of fermions and their superpartners would have to
change in the same way if SUSY was unbroken, so that their contributions in eq. (\ref{VacBubble}) would always cancel. 
However, this is not the case below the SUSY breaking scale.

\subsection*{Bounds on Quintessence Couplings}

\begin{figure}[ht]
\includegraphics[clip,width=1\columnwidth,keepaspectratio]{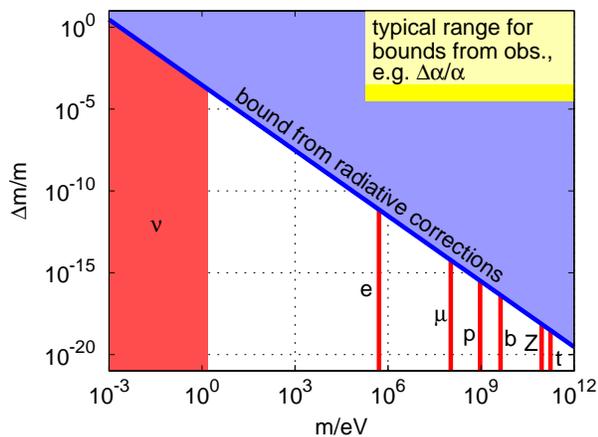}
\caption{\label{CosMassVar}Bounds for cosmic mass variation since $z\sim 2$ from
the radiative correction to the quintessence potential in dependence
of the mass $m$. The red (vertical) lines mark the masses of some standard model particles.
The limits inferred from observations e.g. of $\Delta\alpha/\alpha$ strongly depend
on the considered particle type and further assumptions,
but typically lie around $10^{-4}$ to $10^{-5}$\cite{Uzan:2002vq}.}
\end{figure}

The upper bound (\ref{BoundOnMassVar}) can be directly related to
upper bounds e.g. for the coupling strength to a long range force mediated
by the light scalar field, and for cosmic mass variation. The relative change
of the mass $m_j$ since redshift $z$
can be related to the derivative $d\ln m_j^2/d\ln V''$ using eq. (\ref{MassRedshiftRel}),
\begin{equation}\label{MassVar}
\frac{\Delta m_j}{m_j} \approx \frac{d\ln m_j^2}{d\ln V''} \ln\frac{V''(\phi)}{V''(\phi_0)} \lesssim 3\ln (1+z)\frac{d\ln m_j^2}{d\ln V''},
\end{equation}
which means the bound (\ref{BoundOnMassVar2}) directly gives an upper limit
for the relative mass variation of species $j$ since redshift $z$.
For example, for the variation of the electron mass since $z\sim 2$ we find
\begin{equation}
\frac{\Delta m_e}{m_e} \ll 0.7\cdot 10^{-11}\left(\frac{1-\omega_{de}}{2}\frac{\Omega_{de}}{0.7}\right)^{\frac{1}{3}} ,
\end{equation}
which is at least six orders of magnitude below observational constraints
for a change in the electron-proton mass ratio \cite{Uzan:2002vq}.
For heavier particles, the bounds are even stronger by a factor $(m_e/m)^{4/3}$, see figure \ref{CosMassVar},
e.g. of the order $\Delta m_p/m_p \ll 10^{-15}$ for the proton.
The only known particles which could have a sizeable mass variation due to the bound (\ref{BoundOnMassVar2}) are
neutrinos. Thus models considering mass-varying neutrinos and/or a connection between dark energy and neutrinos 
(see e.g. \cite{Fardon:2003eh,Brookfield:2005td})
do not seem to be disfavored when considering quantum fluctuations. If the bound (\ref{BoundOnMassVar2}) is saturated,
it is even possible that backreaction effects could influence the quintessence dynamics in the recent past,
where the turnover to a dark energy dominated cosmos occurs.

If we consider fermion masses which depend on the quintessence field,
the corresponding coupling mediates a Yukawa-like interaction with
typical range $\sim V''^{-1/2} \sim H_0^{-1}$ and Yukawa coupling strength $y_j\equiv dm_j(\phi)/d\phi$ \cite{Ratra:1987rm}
of species $j$ to this fifth force. 
Inside the horizon, this is a long-range interaction like gravity, which could be detected via a violation
of the equivalence principle. For nucleons, this puts strong bounds on the coupling $y_{p,n}$ of order $10^{-24}$ \cite{Ratra:1987rm}.
On the other hand, 
the  coupling strength is constrained by the bound (\ref{BoundOnMassVar}) via the relation
\begin{equation}
y_j\equiv\frac{dm_j}{d\phi}=\frac{1}{2}m_j\frac{V'''}{V''}\frac{d\ln m_j^2}{d\ln V''}\equiv\frac{m_j}{2M}\frac{d\ln m_j^2}{d\ln V''},
\end{equation}
where we introduced the scale height $M\equiv(d\ln V''/d\phi)^{-1}$ of the quintessence mass which is typically
of the order of the Planck scale \cite{Steinhardt:1999nw}.
If we consider the proton and neutron as effectively massive degrees of freedom,
we obtain an upper limit from the requirement (\ref{BoundOnMassVar2})
\begin{equation}\label{BoundOnProtonYukawaCoupling}
y_{p,n} \ll 0.4\cdot 10^{-35} \left(\frac{M_{pl}}{M}\right) \left(\frac{1{\rm GeV}}{m_{p,n}}\right)^{\frac{1}{3}} 
\left(\frac{1-\omega_{de}}{2}\frac{\Omega_{de}}{0.7}\right)^{\frac{1}{3}},
\end{equation}
which is more than ten orders of magnitude below the limit from the tests of the equivalence principle \cite{Ratra:1987rm},
\begin{equation}
y_{p,n}  \ll 10^{-24}.
\end{equation}
These limits can be compared to the 
corresponding gravitational coupling $\sim m_j/M_{pl}$, e.g. of the order $\sim 10^{-19}$
for the nucleons. Thus the bound in eq. (\ref{BoundOnMassVar2}) also directly gives
a bound for the {\it relative} suppression 
\begin{equation}
\beta_j \equiv \frac{y_j}{m_j/M_{pl}} = \frac{d\ln m_j}{d(\phi/M_{pl})}
\end{equation}
of the coupling strength to the fifth force mediated
by the quintessence field compared to the gravitational coupling, giving roughly
(for $M\sim M_{pl}$, $\omega_{de}+1\lesssim 1$, $\Omega_{de}\sim 0.7$)
\begin{equation}\label{RelStrength}
\beta_j \lesssim \frac{\Delta m_j}{m_j} \ll 4\left(\frac{{\rm meV}}{m_j}\right)^{4/3} \sim 10^{-11} \left(\frac{m_e}{m_j}\right)^{4/3}  .
\end{equation}

Note that the bound from eq. (\ref{BoundOnProtonYukawaCoupling}) also holds for other species (with a scaling $\sim m^{-1/3}$ with mass),
whose quintessence couplings are in general {\it not} constrained by the tests of the equivalence principle \cite{Ratra:1987rm}.
This is also true for dark matter, if it consists of a new heavy species like e.g. a WIMP, 
which severely constrains any coupling via a $\phi$-dependent mass, 
\begin{equation}
y_{dm}=dm_{dm}/d\phi \ll 10^{-36}\, ({\rm TeV}/m_{dm})^{1/3},
\end{equation}
corresponding to a limit of the order 
\begin{equation}
\Delta m_{dm}/m_{dm}\ll 10^{-19}\, ({\rm TeV}/m_{dm})^{4/3}
\end{equation} for a mass variation
between $z\sim 2$ and now from eq. (\ref{RelStrength}).
However, this constraint is not applicable if dark matter is
for example itself given by a scalar condensate, e.g. as in axion models.

%%%%%%%%%%%%%%%%%%%%%%%%%%%%%%%%%%%%%%%%%%%%%%%%%%%%%%%%%%%%%%%%%%%%%%%%%%%%%%%%%%%%%%%%%%%%%%%%%%%%%%%%%%%%%%%%%%%%%%%%%%%%%%%%%%%%%%

\section{\label{sec:3}Self-interaction of the scalar}

%%%%%%%%%%%%%%%%%%%%%%%%%%%%%%%%%%%%%%%%%%%%%%%%%%%%%%%%%%%%%%%%%%%%%%%%%%%%%%%%%%%%%%%%%%%%%%%%%%%%%%%%%%%%%%%%%%%%%%%%%%%%%%%%%%%%%%

If the light scalar field responsible for dark energy
has itself fluctuations described by quantum field theory,
its self-interactions will also contribute to the effective potential.
Typical potentials
used in the context of quintessence, involving e.g. exponentials 
\cite{Ratra:1987rm,Wetterich:1987fm,Albrecht:1999rm,Amendola:1999er}, 
contain self-couplings with an arbitrary number of legs, which are
suppressed by a scale $M$, typically of Planck-size. Such couplings could arise e.g. as an effective
theory by integrating out some unknown high-energy degrees of freedom.
Usually, quantum fluctuations in the presence of
such couplings can be treated by an expansion in the inverse of the suppression scale $M$.
However, in the case of a scalar field, the high-energy sector does not completely decouple 
due to the well-known quadratically divergent contributions. In the context of an effective theory, 
the quadratically divergent diagrams, e.g. the tadpole graph,
are intrinsically governed by a scale $\Lambda$ which is characteristic for
the high-energy scale up to which the effective theory is valid. In the simplest case,
$\Lambda$ can be imagined as a cutoff for the momentum cycling in the loop.
Both high-energy scales $M$ and $\Lambda$
could of course be related in a way depending on the unknown underlying high-energy theory.
Since the suppression scale $M$ could be as large as the Planck scale,
it is even possible that the same is true for $\Lambda$.
However, since unknown quantum gravity effects will play an important role in this regime,
we will just assume an upper bound $\Lambda\lesssim M_{pl}$.

In order to establish a meaningful approximation, it would be desirable
to resum all contributions proportional to powers of $\Lambda/M$,
whereas the tiny mass of the quintessence field given by $V''$,
which is typically of the order of the Hubble scale today,
could admit a perturbative expansion e.g. in powers of $V''/\Lambda^2$.
In the following we will motivate that such an expansion might indeed be possible,
and calculate the leading contributions explicitely.

A typical feature of quintessence potentials
is that the self-couplings $V^{(k)}(\phi)$ with $k$ lines are suppressed like 
\begin{equation}\label{PowCountCoupling}
V^{(k)}(\phi) \sim \mathcal{O}(V''/M^{k-2})
\end{equation}
with $M\equiv(d\ln V''/d\phi)^{-1}$, 
where  for example $M\sim M_{pl}$
for exponential potentials and $M\sim\phi$ for inverse power law and tracking potentials \cite{Steinhardt:1999nw}
with $\phi\sim M_{pl}$ in the present epoch. 
The effective potential can be calculated by the sum over all 1PI vacuum diagrams with
propagator $G(p)=i/(p^2- V''(\phi))$ and vertices $-iV^{(k)}(\phi)$ with $k$ legs. 
If one uses the power-counting estimate in eq. (\ref{PowCountCoupling}),
then a $L$-loop diagram with
$\mathcal{V}_k$ vertices with $k\geq 3$ legs and $E$ external lines  
is (in terms of dimension-full quantities, i.e. disregarding logarithms etc.) proportional to
\begin{equation}
\prod_{k}\left(V^{(k)}\right)^{\mathcal{V}_k}\Lambda^{2(L-\mathcal{V}+1)} 
\sim \left(\frac{V''}{\Lambda^2}\right)^{\mathcal{V}}\left(\frac{\Lambda^2}{M^2}\right)^{L+1}M^{4-E},
\end{equation} 
which shows that the diagrams with only one vertex are the leading contribution in $V''/\Lambda^2 \ll 1$,
whereas there is not necessarily a strong suppression of
contributions with high $L$.
This indicates that in the case of quintessence-like potentials
an appropriate expansion parameter is given by the number of vertices $\mathcal{V}$,
whereas the loop expansion becomes meaningless if $\Lambda\gtrsim M$.

\subsection*{Bubble Approximation}

We will now calculate the effective potential in leading order
in the number of vertices $\mathcal{V}$ and show that it can be consistently renormalized
for quintessence potentials obeying eq. (\ref{PowCountCoupling}), up to higher order corrections.
The graphs with $\mathcal{V}=1$ are ``multi-bubble" graphs, i. e.
graphs where an
arbitrary number of tadpoles is attached to the vertex.
Thus, diagrammatically, the effective mass, i.e. the second derivative of the
effective potential\footnote{It is also possible but less convenient to
calculate $V_{\textit{eff}}$ directly.}, is given in leading order in $\mathcal{V}$
by the infinite sum
\begin{equation}
\begin{array}{lcl}\label{TadSum}
\displaystyle V_{\textit{eff}}''(\phi) 
&\displaystyle=&\displaystyle  
V''(\phi) 
+ \rule{0mm}{7mm}\parbox{11mm}{\includegraphics{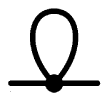}} 
+ \parbox{11mm}{\includegraphics{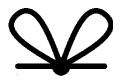}}
+ \parbox{11mm}{\includegraphics{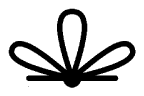}}
+ \dots \\
&\displaystyle=&\displaystyle 
\sum_{L=0}^\infty \frac{V^{(2L+2)}(\phi)}{2^LL!}\left(\int\!\!\!\frac{d^4q}{(2\pi)^4}\frac{1}{q^2+V''(\phi)}\right)^L \\
&\displaystyle=&\displaystyle
\exp\left(\frac{1}{2}I_1(m^2)\frac{d^2}{d\phi^2}\right) V''(\phi)\, \rule[-6mm]{0.15mm}{10mm}_{\, m^2=V''(\phi)} ,
\end{array}
\end{equation}
with euclidean momenta and $I_1$ as defined in (\ref{Ik}). The last line is a compact notation
where the derivatives in the exponential act on the tree level mass $V''(\phi)$ on the right hand side.
Note that this is the infinite resummation 
referred to above, in leading order in  $\mathcal{V}$.
The physically relevant scale of the quadratically divergent tadpole integral $I_1$ is given by a high energy scale
characteristic for the effective theory, as discussed above. To implement this behaviour in a consistent way,
we split
\begin{equation}
I_1 \equiv I_1^{\rm finite} + I_1^{\rm div},
\end{equation}
where the finite value should represent the physically relevant part. 
The formally divergent part can be consistently reabsorbed
by shifting the potential by an analytic function of the field value $\phi$,
\begin{equation}
V(\phi)\rightarrow V(\phi)+\delta V(\phi),
\end{equation}
as is shown in detail in appendix \ref{app:Ren},
up to corrections which are suppressed by a factor of the order $V''/I_1^{\rm finite}$
relative to the leading contributions\footnote{In order to renormalize also
the sub-leading contributions, it would be necessary to include also two-vertex graphs since they are of the
same order.} and assuming eq. (\ref{PowCountCoupling}) for the potential.
Thus the bubble approximation allows a self-consistent renormalization
in leading order in $V''/I_1^{\rm finite}$, with the important result
(see appendix \ref{app:Ren})
\begin{equation}\label{TadSumRenorm}
V_{\textit{eff}}''(\phi) = 
\exp\left(\frac{1}{2}I_1^{\rm finite}(m^2)\frac{d^2}{d\phi^2}\right) V''(\phi)\, \rule[-6mm]{0.15mm}{10mm}_{\, m^2=V''(\phi)}.
\end{equation}
up to terms which are relatively suppressed by $V''/I_1^{\rm finite}$.
This result means that the $L$-loop contribution to $V_{\textit{eff}}''(\phi)$, see eq. (\ref{TadSum}),
which is proportional to $(I_1)^L$, leads to a leading-order contribution of the form $(I_1^{\rm finite})^L$
after renormalization, i.e. ``renormalization and raising to a power do  commute" in this case.

The physically relevant part $I_1^{\rm finite}$ will be determined by an
underlying theory, possibly involving quantum gravity effects,
which produces the quintessence potential as effective
theory. In the present approach we parameterize this
unknown function by a Taylor expansion in $V''$,
\begin{equation}\label{FiniteTadpole}
I_1^{\rm finite}(V'') \ \equiv \ \pm\frac{\mu^2}{16\pi^2} \,+\, \dots\, ,
\end{equation}
where the dots include linear and higher terms in $V''$.
Since $V''$ is typically of the order of the Hubble scale,
we neglect these contributions with respect to the
high-energy scale $\mu\lesssim M_{pl}$,
which should maximally be of the order of the scale $\Lambda$
introduced in the first part of this section. 
Furthermore, one cannot exclude \textit{a priori} the possibility
that $I_1^{\rm finite}$ can
be negative or positive, indicated by the two signs.
This again depends on the embedding into the underlying theory,
where unknown quantum gravity effects could play a major role.
Additionally, there are examples like the Casimir effect,
where it is known that the sign of the renormalized $00$-component of the
energy-momentum tensor can be negative or positive, depending
e.g. on boundary conditions and geometry, even though the unrenormalized
contribution is positive definite.

Altogether, the leading contribution to the effective potential
is given by 
\begin{equation}\label{EffPot}
V_{\textit{eff}}(\phi)= \exp\left(\pm\frac{1}{32\pi^2}\mu^2\frac{d^2}{d\phi^2}\right) V(\phi),
\end{equation} 
up to corrections suppressed by $V''/\mu^2\ll 1$.
This is the main result of this section.
It has been obtained by integrating eq. (\ref{TadSumRenorm}) twice with respect to $\phi$
after using eq. (\ref{FiniteTadpole}) and absorbing the constant of integration, 
which corresponds to the
quartic divergence, into $\delta V$. Thus the potential is, as usual,
only determined up to an arbitrary additive constant, which we set to zero,
corresponding to the unresolved ``cosmological constant problem".

\subsection*{Stability of Quintessence Potentials}

The above result (\ref{EffPot}) gives a simple prescription to estimate
the stability of a quintessence potential $V(\phi)$ under its self-interactions.
In the following, we will investigate the effect of an operator of the form
\begin{equation}\label{DerivOp}
\exp\left(\pm\frac{\mu^2}{32\pi^2}\frac{d^2}{d\phi^2}\right)
\end{equation}
on some typical potentials often used in dynamical dark energy scenarios.
One archetype class of potentials are given by
(combinations of \cite{Barreiro:1999zs,Neupane:2003cs}) exponential potentials \cite{Ratra:1987rm,Wetterich:1987fm,Amendola:1999er}.
Remarkably, an exponential of the field $\phi$ is form-invariant under the action
of the operator (\ref{DerivOp}). Consider e.g. the following finite or infinite sum
of exponentials,
\begin{equation}\label{ExpPot}
V(\phi) = \sum_jV_j\exp\left(-\lambda_j\frac{\phi}{M_{pl}}\right).
\end{equation}
The only effect of  (\ref{DerivOp}) is a simple rescaling of the prefactors $V_j$ according to
\begin{equation}
V_j \ \rightarrow \ V_j\exp\left(\pm\frac{\lambda_j^2\mu^2}{32\pi^2M_{pl}^2}\right).
\end{equation}
This extends the result of ref. \cite{Doran:2002bc} for the one-loop case,
which would corresponds to the first term in a Taylor expansion of (\ref{DerivOp}). 
Note that if $\mu\sim M_{pl}$ 
the correction can be of an important size, and can influence the relative strength
of the exponentials in (\ref{ExpPot}). 
The necessary condition of validity $V''(\phi)\ll\mu^2$ for the
bubble approximation is typically fulfilled when $V\ll \mu^2M_{pl}^2$,
which implies that it is applicable if $\mu\gg H_{\textit{max}}$, where $H_{\textit{max}}$ is the maximum value
of the Hubble parameter where the field $\phi$ plays a role.
For example, $H_{\textit{max}}$ could be the inflationary scale $H_{\textit{inf}}$, e. g. around $10^{13}{\rm GeV}$.
Altogether, exponentials seem to be
stable under the considered radiative corrections.

\begin{figure}[ht]
\includegraphics[clip,width=1\columnwidth,keepaspectratio]{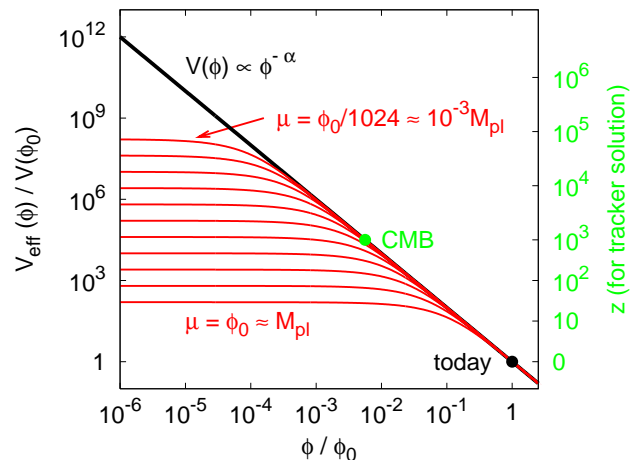}
\caption{\label{InvPow}Effective potential $V_{\textit{eff}}(\phi)$ for an inverse power law potential
$V(\phi)\propto\phi^{-\alpha}$ with $\alpha=2$ for various values
of $\mu$ on a double logarithmic scale and normalized
to todays value $V(\phi_0)$. From top to bottom,
$\mu$ is enlarged by a factor $2$ for each red (thin) line.
The black (thick) line is the tree potential $V(\phi)$. 
For $\phi\lesssim\mu/(4\pi\sqrt{2})$ the effective potential approaches a constant value,
whereas $V$ gets huge. The redshift-scale
on the right-hand side applies for the tracking solution of $V$ {\it only} and illustrates
when the deviations of $V_{\textit{eff}}$ from $V$ become relevant in cosmic history going backward
from $\phi/\phi_0=1$ (today).}
\end{figure}

Another often discussed class of potentials are (combinations) of inverse powers of the
field $\phi$ \cite{Ratra:1987rm,Steinhardt:1999nw,Brax:1999yv,Doran:2002bc},
\begin{equation}\label{InvPot}
V(\phi) = \sum_\alpha c_\alpha\phi^{-\alpha}.
\end{equation}
The action of the operator (\ref{DerivOp}) yields
\begin{equation}\label{InvEffPot}
\begin{array}{lcl}
\displaystyle V_{\textit{eff}}(\phi) \!\!
&\displaystyle= &\displaystyle \!
\sum_\alpha\frac{c_\alpha\phi^{-\alpha}}{\Gamma(\alpha)}\sum_{L=0}^\infty\frac{\Gamma(\alpha+2L)}{L!} 
\left(\frac{\pm\mu^2}{32\pi^2\phi^2}\right)^L \\
&\displaystyle= &\displaystyle \!
\sum_\alpha\frac{c_\alpha\phi^{-\alpha}}{\Gamma(\alpha)}\int_{0}^\infty\!\!\!dtt^{\alpha-1}
\exp\left(-t\pm\frac{\mu^2}{32\pi^2\phi^2}t^2\right), 
\end{array}
\end{equation}
where the $\Gamma$-function inside the sum over $L$ has been replaced by its definition via an
integration over the positive real axis in the second line. This integral only gives a
finite result if the negative sign in the exponent is used, which we will
therefore assume from now on.
We will first discuss two limiting cases where the integral can be easily solved analytically.
For large field values $\phi\gg\mu$, which corresponds to small potential energy and
-curvature, the second term in the exponent $-t-\mu^2/(32\pi^2\phi^2)t^2$ appearing
in (\ref{InvEffPot}) can be neglected, which implies that asymptotically
\begin{equation}\label{EffPotForLargePhi}
V_{\textit{eff}}(\phi) \rightarrow V(\phi) \equiv \sum_\alpha c_\alpha\phi^{-\alpha}, \quad \phi\rightarrow\infty.
\end{equation}
This means the low energy regime where $V$ and its derivatives go to zero is
not changed. For the opposite limit where $\phi\ll\mu$,
the integral in the last line of (\ref{InvEffPot}) can be calculated by neglecting the
first term in the argument of the exponential, 
\begin{equation}
\begin{array}{lcl}
\displaystyle V_{\textit{eff}}(\phi) &\displaystyle \rightarrow &\displaystyle
\sum_\alpha \frac{c_\alpha\phi^{-\alpha}}{\Gamma(\alpha)} 
\frac{1}{2}\Gamma(\frac{\alpha}{2}) \left(\frac{\mu^2}{32\pi^2\phi^2}\right)^{-\frac{\alpha}{2}} \\
&\displaystyle = &\displaystyle 
\sum_\alpha \frac{\Gamma(\frac{\alpha}{2})}{2\Gamma(\alpha)} c_\alpha\left(\frac{\mu}{4\pi\sqrt{2}}\right)^{-\alpha} = {\rm const}.
\end{array}
\end{equation}
Thus the effective potential approaches a
constant finite value for $\phi\lesssim\mu/(4\pi\sqrt{2})$ of the order $V(\mu)$, 
which gets smaller for larger values of $\mu$, see figure \ref{InvPow} for the special case $V\propto\phi^{-2}$. 
Furthermore, it is easy to see that also $V_{\textit{eff}}''(\phi)$
approaches a constant value 
\begin{equation}\label{EffMassForSmallPhi}
V_{\textit{eff}}''(\phi)\rightarrow \sum_\alpha \frac{\Gamma(\frac{\alpha+2}{2})}{2\Gamma(\alpha)} c_\alpha\left(\frac{\mu}{4\pi\sqrt{2}}\right)^{-(\alpha+2)}.
\end{equation}
These results show that the singular behaviour of the
potential $V(\phi)$, see eq. (\ref{InvPot}), for $\phi\rightarrow 0$ is not present in the effective potential, where
a constant value of the order $V(\mu)$ is approached instead. 

The bubble approximation requires that $V''\ll\mu^2$,
which means we again find that the approximation is valid as long as
$\mu\gg H_{\textit{max}}$, with $H_{\textit{max}}\lesssim H_{\textit{inf}}$, 
as for the exponential potential, if we simply assume tracking behaviour \cite{Steinhardt:1999nw}.
If the scale $\mu$ is very large, the bubble approximation is always valid
in the range of cosmological interest anyway. For example, for\footnote{
For $V=c_\alpha\phi^{-\alpha}$ typical values within quintessence scenarios are $c_\alpha\sim H_0^2M_{pl}^{\alpha+2}$, 
e.g. $c_2\sim (100{\rm MeV})^6$ \cite{Steinhardt:1999nw}.} $V\propto\phi^{-2}$
as in figure \ref{InvPow}, one finds that the condition $V''\ll\mu^2$ with $\mu\gtrsim 10^{-3}M_{pl}$
is always fulfilled as long as $\phi/\phi_0>10^{-16}$, where $\phi_0\sim M_{pl}$
is the field value today, which is far below the relevant range of $\phi$.
Furthermore, eq. (\ref{EffMassForSmallPhi}) shows that due to the
absence of a singular behaviour in the second derivative of the {\it effective} potential the condition
$V_{\textit{eff}}''/\mu^2\ll 1$ can hold for {\it all} positive values of $\phi$ if 
$\mu\gg \max_\alpha(c_\alpha)^{1/(\alpha+4)}$. This might indicate that the bubble approximation
is indeed applicable in the {\it total} range of field values. However, to show this formally it is
necessary to additionally perform a 2PI approximation \cite{Cornwall:1974vz}
where a dressed propagator containing $V_{\textit{eff}}''$ instead of $V''$ appears in the loops.
This is left to future work.

\begin{figure}[ht]
\includegraphics[clip,width=1\columnwidth,keepaspectratio]{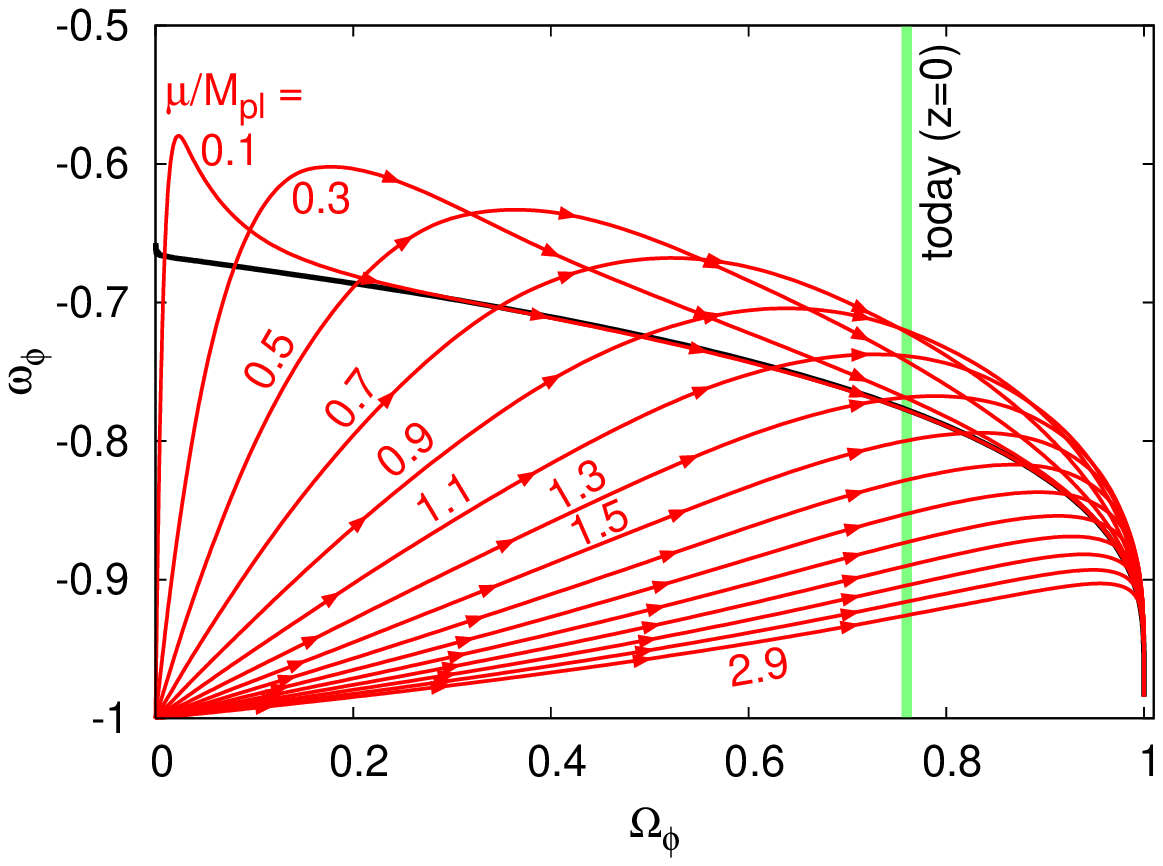}
\includegraphics[clip,width=1\columnwidth,keepaspectratio]{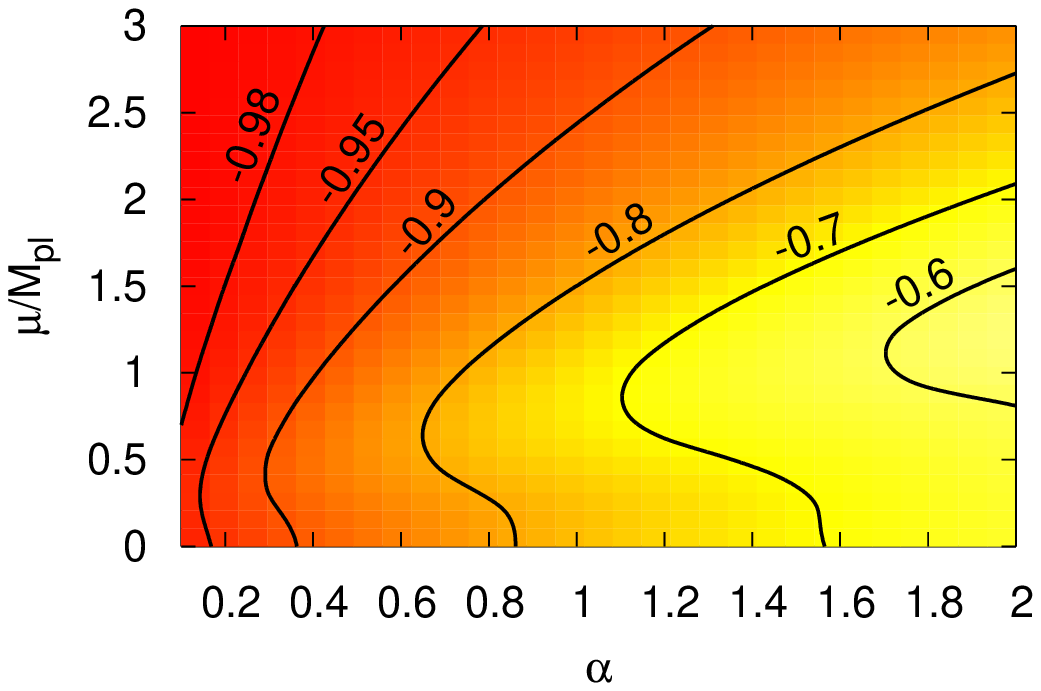}
\caption{\label{omegaOmega}Top: Evolution in the $(\Omega_\phi,\omega_\phi)$-plane for the effective potential
$V_{\textit{eff}}$ of an inverse power law potential
$V\propto \phi^{-\alpha}$ with $\alpha=1$ for various values
of $\mu$ keeping $H_0=73{\rm km}/{\rm s} {\rm Mpc}$ and $\Omega_{de}\equiv\Omega_\phi(z=0)=0.76$ fixed. From top to bottom,
$\mu$ is enlarged by $0.2 M_{pl}$ for each red (thin) line starting from $\mu=0.1 M_{pl}$.
The black (thick) line is the tracking solution in the tree potential $V(\phi)$, from
which  the solutions deviate considerably for large $\mu$. The four arrows on each trajectory
mark the points with redshifts $z=2,1,0.5,0.1$ from left to right.\\
Bottom: Contour plot of the equation of state $\omega_{de}$ today $(z=0)$ 
using the effective potential $V_{\textit{eff}}$ of a potential
$V\propto \phi^{-\alpha}$ depending on $\mu$ and $\alpha$. $\mu=0$ corresponds to the case $V_{\textit{eff}}\equiv V$.
Again, we fixed $H_0$ and $\Omega_{de}=0.76$ as above.}
\end{figure}

Let us now estimate in how far 
typical tracking quintessence models are changed by considering
the effective potential from eq. (\ref{InvEffPot}).
Since the field value today is typically of the order of the Planck scale \cite{Steinhardt:1999nw},
the large-field limit eq. (\ref{EffPotForLargePhi}), where the effective potential
approaches the tree level potential and the corrections are negligible, is only applicable when $\mu\lll M_{pl}$. 
For values up to $\mu\lesssim M_{pl}/10$ the field $\phi$
can have a tracking solution. 
The redshift $z_{\rm quant}$ in cosmic history where the effective potential
starts to deviate from the tracking potential, see figure \ref{InvPow}, gives a rough
estimate at which redshift the tracking sets in. For a potential
dominated by a single inverse power $V\propto\phi^{-\alpha}$ 
we obtain, requiring a deviation of the effective potential of less than $1\%$ and
using the tracking solution during matter and radiation domination with equation of state 
$\omega_{\phi}=\frac{\alpha}{\alpha+2}(1+\omega_B)-1$ \cite{Steinhardt:1999nw}, with $\omega_B=0,1/3$ 
respectively,
\begin{equation}
z_{\rm quant} \sim \left(\frac{M_{pl}/10}{\mu/(4\pi\sqrt{2\alpha(\alpha+1)})}\right)^{\frac{\alpha+2}{3(1+\omega_B)}},
\end{equation}
e.g. assuming $\mu\sim M_{pl}/100$ ($M_{pl}\equiv 1/\sqrt{G}$) one gets $z_{\rm quant}\sim 300$ for $\alpha=2$ 
and $z_{\rm quant}\sim 130$ for $\alpha=1$. Similar bounds should also
hold for other types of potentials, e.g. like the SUGRA-potential \cite{Brax:1999yv},
which are dominated by an inverse-power law behaviour at redshifts $z\gg 0.5$.
For values $\mu\gtrsim M_{pl}/10$, there can be large deviations from
the tracking solution even at low
redshifts and today, see figure \ref{omegaOmega} for 
an exemplary case with $V\propto \phi^{-\alpha}$. 
If $\mu$ is extremely large, there is a
direct transition from the slow roll regime with $\phi\lesssim\mu$, equation of state $\omega_\phi\sim -1$ and
dark energy fraction $\Omega_\phi\lll 1$ in the flattened effective potential  $V_{\textit{eff}}$
to the Dark Energy dominated accelerating solution for $\phi\gtrsim M_{pl}$ with $\Omega_\phi\rightarrow 1$
and $\omega_\phi\rightarrow -1$, and thus the solution
never performs scaling with $\omega_\phi=-\frac{2}{\alpha+2}$ as for $V$. In the case $\alpha=1$,
the equation of state today $\omega_{de}\equiv\omega_\phi(z=0)$ is enhanced for $0.1\lesssim\mu/M_{pl}\lesssim 1.3$
compared to the tracking value, and gets smaller for even larger\footnote{Note that even when $\mu\gtrsim M_{pl}$
the renormalized tadpole $I_1^{\rm finite}$ can still be sub-Planckian due to the loop factor $1/16\pi^2$.} 
$\mu$, see figure \ref{omegaOmega}. Moreover, the sign of $d\omega_\phi/dz$ can change depending on $\mu$.

%%%%%%%%%%%%%%%%%%%%%%%%%%%%%%%%%%%%%%%%%%%%%%%%%%%%%%%%%%%%%%%%%%%%%%%%%%%%%%%%%%%%%%%%%%%%%%%%%%%%%%%%%%%%%%%%%%%%%%%%%%%%%%%%%%%%%%

\section{\label{sec:4}Effective Action and Curved Background}

%%%%%%%%%%%%%%%%%%%%%%%%%%%%%%%%%%%%%%%%%%%%%%%%%%%%%%%%%%%%%%%%%%%%%%%%%%%%%%%%%%%%%%%%%%%%%%%%%%%%%%%%%%%%%%%%%%%%%%%%%%%%%%%%%%%%%%

Since any dynamical dark energy scenario is generically used in a curved space-time setting,
e. g. described by a Robertson-Walker metric, it is important to study also the quantum
fluctuation on such a background. Generically, dynamical dark energy scenarios
making use of a scalar field involve non-renormalizable interactions suppressed by some high-energy scale
up to the Planck scale. Therefore it is important to include such interactions when discussing
quantum induced non-minimal couplings between the dark energy scalar field and gravity. 

In this section, we will investigate 
the behaviour of a scalar field whose potential contains {\it at least} one term which is
non-renormalizable in the usual power-counting sense, e. g. $V(\phi)\ni \phi^n$ with $n>4$,
using the semiclassical treatment of curved space-time in quasi-local approximation.
This includes most quintessence models as well as models that can be
reformulated using a canonical scalar field.

On a Minkowski background, the one-loop contribution to the effective potential
requires the introduction of counterterms proportional to $V''(\phi)$ and $V''(\phi)^2$.
If $V(\phi)$ is an analytic function of $\phi$ and contains at least one power of the
field larger than $4$, this immediately implies that in order to remove all divergences
it is necessary that $V(\phi)$ contains terms with {\it all} powers of $\phi$. In other words,
the structure of the one-loop quantum corrections enforces to keep either only
renormalizable terms in $V(\phi)$ or to admit an arbitrary analytic form for $V(\phi)$.

In the case of a renormalizable potential in curved space-time
it is well known \cite{Birrell:1982ix,Elizalde:1994ds,Hu:1984js} that
in order to be able to remove all divergences at the one-loop level it is necessary to include
also non-minimal coupling terms with the curvature scalar $R$ of the form $\phi^2R$
as well as terms proportional to $R$, $R^2$, the square of the Weyl tensor
$C = R_{\mu\nu\rho\sigma}R^{\mu\nu\rho\sigma}-2R_{\mu\nu}R^{\mu\nu}+\frac{1}{3}R^2$,
the Gauss-Bonnet invariant $G = R_{\mu\nu\rho\sigma}R^{\mu\nu\rho\sigma}-4R_{\mu\nu}R^{\mu\nu}+R^2$,
$\Box R$ and $\Box\phi^2$, where $\Box$ is the covariant D'Alembertian. The latter three terms are total derivatives and thus not
relevant for the dynamics, but they are needed for the cancellation of divergences
and do appear in the dynamics if their running is considered  \cite{Elizalde:1994ds}.

The question we want to address here is how this set of operators is enlarged
when also non-renormalizable terms appear. This means we want to find an (infinite)
set of operators that closes under one-loop quantum corrections, i.e. any divergence produced by these operators
can be reabsorbed into one of them. Therefore we use a quasi-local approximation accessible
through the heat kernel expansion in the form suggested by \cite{Jack:1985mw}.
The outcome of this approach is that, in addition to the operators $C$ and $G$ needed in the
renormalizable case, it is necessary to use a generalized potential $V(\phi,R)$
that is an arbitrary analytic function of the field $\phi$ {\it and of the curvature scalar} $R$
and furthermore to include total derivative terms of the form $\Box B(\phi,R)$ where
$B(\phi,R)$ is also an arbitrary analytic function of $\phi$ and $R$. In other words,
the ``smallest" Lagrangian which is stable under one-loop corrections in curved space
and includes non-renormalizable interactions is of the form
\begin{equation}\label{LagrangeCurved}
\mathcal{L} = \frac{1}{2}(\partial\phi)^2-V(\phi,R)+\epsilon_1C+\epsilon_2G+\Box B(\phi,R),
\end{equation}
with dimensionless coupling constants $\epsilon_1$ and $\epsilon_2$,
and admitting arbitrarily high powers of {\it both} $\phi$ and $R$ to be contained in $V$ and $B$.

We will now show this result. In Appendix \ref{app:HKE},
where we also calculate the one-loop effective action explicitely using zeta-function
regularization,
we show that apart from counterterms proportional to $C$ and $G$ one has to
introduce a counterterm proportional to 
\begin{equation}\label{CurvCounter}
\left(\frac{\partial^2V}{\partial\phi^2}-\frac{R}{6}\right)^2,
\end{equation}
i.e. all
operators contained in this expression have to be present in the tree level Lagrangian.
Now assume for the moment that $V$ is an (arbitrary) analytic function $f_0(\phi)$ {\it only} of $\phi$, i.e.
can be written as a series in $\phi^n$ with $n\geq 0$, as required in flat space. The Lagrangian then
has to include all terms contained in $(f_0''-R/6)^2$, especially those proportional to $\phi^{n-2}R$ for all $n\geq 2$.
Consequently, one has to add a term of the form $f_1(\phi)R$ to the Lagrangian with an (arbitrary) analytic function $f_1(\phi)$,
giving the counterterm (\ref{CurvCounter}) the form $(f_0''+f_1''R-R/6)^2$. This implies
the presence of $\phi^{n-2}R^2$-terms for all $n\geq 2$ in the Lagrangian, i.e. a term $f_2(\phi)R^2$ also
has to be added, and so on. Recursively, this implies that $V=\sum_n f_n(\phi)R^n$, i.e. $V$ is an
arbitrary analytic function of $\phi$ {\it and} $R$, as was to be shown. 
Furthermore,  
counterterms proportional to $\Box(\partial^2V(\phi,R)/\partial\phi^2-R/5)$ (see appendix  \ref{app:HKE})
immediately imply that also $B=B(\phi,R)$.

Some comments are in order:\\
(i) The upper result can also be rephrased by stating that even if one starts with a tree level
Lagrangian where the scalar is minimally coupled to gravity at a certain scale,
quantum fluctuations will induce non-minimal coupling terms with arbitrarily high powers of $R$
in the effective action through the running of the corresponding couplings.
\\ 
(ii) There are two special cases where the introduction of arbitrarily high powers of $R$
can be avoided in a way which is stable under one-loop corrections: First, as expected, the renormalizable case,
where for the choice $V(\phi,R)=\Lambda+m^2\phi^2/2+\lambda\phi^4+\xi\phi^2R/2+M^2R+\epsilon_0R^2$ 
and $B(\phi,R)=\epsilon_3 R+\eta \phi^2$ all divergences can be absorbed, as already mentioned above.
Second, the conformally coupled case, where $V(\phi,R)=V(\phi)+\xi\phi^2R/2$ with $\xi=1/6$, and $B(\phi,R)=\epsilon_3R+B(\phi)$. 
In this
case, all divergences proportional to higher powers of $R$ vanish, since $R$ is canceled
in eq. (\ref{CurvCounter}). 
In other words, if we write $V=\sum_n\sum_mc_{nm}\phi^nR^m$,
there is a fixed point where all
couplings $c_{nm}$ for $m\geq 1$ do not run for $c_{21}=\xi/2=1/12$, $c_{nm}=0$ for $n>2,m\geq 1$ and arbitrary\footnote{
Since these couplings do not receive any quantum corrections in the conformally coupled case,
they could be set to zero by hand.} $c_{1m}$
and $c_{0m}$ for $m\geq 1$. However, within many typical quintessence scenarios
where the field value $\phi$ is of the order of the Planck scale today \cite{Steinhardt:1999nw},
a non-minimal coupling of the form $\xi\phi^2R/2$ is restricted since it leads e.g.
effectively to a variation of the Newton constant $(16\pi G_{\textit{eff}})^{-1}=(16\pi G)^{-1}-\xi\phi^2/2$.
Limits found by several authors \cite{Chiba:1999wt,Perrotta:1999am} for specific models
lie in the range $|\xi|\lesssim 3\cdot 10^{-2}$ which is far below the conformal coupling.
\\
(iii) It is possible to rewrite the Lagrangian (\ref{LagrangeCurved}) so that
$R$ enters only linearly by performing a
conformal transformation $g_{\mu\nu}\rightarrow e^{-\sigma}g_{\mu\nu}$ to the Einstein frame. However,
this introduces an additional scalar $\sigma$ coupled to $\phi$ into the theory.
Technically, one rewrites the Lagrangian (\ref{LagrangeCurved}) in the form
\begin{equation}
\mathcal{L} = \frac{1}{2}(\partial\phi)^2-V(\phi,A) - B(R-A) +\mathcal{L}_c,
\end{equation}
with auxiliary fields $A$ and $B$, see e.g. \cite{Capozziello:2006dj}. $\mathcal{L}_c$ denotes all additional contributions
including the terms proportional $C$ and $G$. After eliminating $B=\partial V/\partial A$
and performing the conformal transformation with $\sigma=\ln(2\kappa^2\partial V/\partial A)$ ($\kappa^2\equiv 8\pi G$)
the corresponding Lagrangian is
\begin{equation}
\mathcal{L} = \frac{-R}{2\kappa^2} + \frac{3}{2\kappa^2}(\partial\sigma)^2 + e^{-\sigma}\frac{1}{2}(\partial\phi)^2
-U(\phi,\sigma) + e^{-2\sigma}\mathcal{L}_c.
\end{equation}
Apart from the $\sigma$-dependent kinetic term of $\phi$ both fields interact through the 
potential given by $U=(V-A\,\partial V/\partial A)e^{-2\sigma}$, where $A=A(\sigma,\phi)$
is given by the inversion of $\sigma=\ln(2\kappa^2\partial V/\partial A)$ w.r.t. $A$.
However, the physical equivalence of conformally related frames is 
not manifestly obvious (see discussion in \cite{Capozziello:2006dj})
and the calculation of quantum corrections in the different frames can yield
inequivalent results due to a nontrivial Jacobian
in the path integral \cite{Doran:2002bc}. Thus the main clue is that the
non-minimal coupling with higher powers of $R$ of the general form $V(\phi,R)$
cannot be simply rescaled away without profoundly changing the scalar sector of the
Lagrangian.
\\
(iv) Since the potential is non-renormalizable it is necessary to introduce
the infinite set $\phi^nR^m$ (with $n,m\geq 0$) of operators to cancel
the one-loop divergences. This means that the corresponding couplings
are not predicted by the theory, at least at a certain reference scale,
but have to be determined in principle by comparison with experiment. 
Of course, this is far from being possible.
Nevertheless, the result suggests that the framework for searching
for an explanation of cosmic acceleration could be a combination
of the two extreme cases of quintessence models with $V=V(\phi)$ on the one hand and
modified gravity scenarios corresponding to $V\ni f(R)$ (see e.g. 
\cite{Nojiri:2006ri,Capozziello:2003tk,Copeland:2006wr} for reviews)
on the other.

The main result of this section is that whenever one considers a scalar with non-renormalizable
interactions and non-conformal coupling, one-loop quantum fluctuations will induce the presence of terms with arbitrarily high
powers of the curvature scalar $R$ in the action. In the case of dark energy,
this means that the quantum fluctuations in the quintessence field could lead
to a modified gravity which differs considerably from the standard Friedmannian behaviour.
Consequently, in a quantized picture a scalar condensate with non-renormalizable potential accounting for dark energy
goes hand in hand with a modified gravity theory described by
a generalized potential $V(\phi,R)$. In fact, such a potential
could not only give rise to an explanation of the accelerated expansion
due to a quintessence-like behaviour, but also through the modification of gravity.

%%%%%%%%%%%%%%%%%%%%%%%%%%%%%%%%%%%%%%%%%%%%%%%%%%%%%%%%%%%%%%%%%%%%%%%%%%%%%%%%%%%%%%%%%%%%%%%%%%%%%%%%%%%%%%%%%%%%%%%%%%%%%%%%%%%%%%

\section{\label{sec:Conclusions}Summary and Conclusions}

%%%%%%%%%%%%%%%%%%%%%%%%%%%%%%%%%%%%%%%%%%%%%%%%%%%%%%%%%%%%%%%%%%%%%%%%%%%%%%%%%%%%%%%%%%%%%%%%%%%%%%%%%%%%%%%%%%%%%%%%%%%%%%%%%%%%%%

In this work we have investigated the effect of quantum fluctuations
in the context of typical dynamical dark energy scenarios
like quintessence models, where an extremely light rolling scalar field 
supplies the present cosmic acceleration, in some sense similar to the inflaton in the
early universe. 

First couplings between the quintessence field and heavier degrees of freedom,
like the standard model fermions or dark matter, have been discussed.
We constrained the discussion to couplings that can effectively be written as a field-dependent mass term.
These couplings have to be extremely small  even though we fine-tune the energy density, slope and mass
of the quintessence field at its todays value by appropriate renormalization conditions
for the quartic, quadratic and logarithmic divergences in the induced effective potential.
This leads to a bound on time-varying masses between $z\sim 2$ and now of the order $10^{-11}$ for the electron and
scaling proportional $m^{-4/3}$ with mass, assuming the mass variations are not themselves finely tuned
in such a way that the total shift in vacuum energy is negligible.
Moreover, we found that the coupling strength
to a fifth force mediated by the quintessence field has to be suppressed by a number of the same
order relative to its gravitational coupling strength. Only neutrinos could have a large
mass variation and interact with the quintessence field as strong as with gravity. 

Second we introduced a suitable approximation scheme to investigate the impact of quintessence self-couplings
on the {\it shape} of the effective potential, while an undetermined additive constant has
been fine-tuned to be zero, thus bypassing the unresolved ``cosmological constant problem".
We showed that the quantum fluctuations to the scalar potential can be consistently renormalized
in leading order in $V''/\mu^2$, where $\mu$ is a high energy scale characteristic for an
underlying theory and $V''$ the square of the quintessence mass 
assumed to be of the order of the Hubble parameter.
While potentials involving exponentials
just get rescaled, inverse power law potentials are flattened at small field values. 
The effective potential approaches a finite maximal value, 
thus truncating the singular behaviour of the inverse power law in the field range of interest. 
This distortion of the potential
can directly play a role cosmologically if $\mu$ is large, roughly $\mu\gtrsim M_{pl}/10$,
and moreover affect general properties like tracking behaviour.

Third we investigated non-minimal gravitational couplings induced by quantum fluctuations. 
Since quintessence potentials
usually cannot be taken to be a quartic polynomial in the field, non-minimal couplings
apart from a term of the form $\phi^2R$ can be induced. We showed that at one-loop
all couplings of the form $\phi^nR^m$ with integers $n$ and $m$ have to be included,
and will be induced by quantum corrections unless the field is exactly conformally coupled.
Moreover, we showed that this type of non-minimal coupling which is nonlinear in $R$ cannot be
simply removed by a Weyl rescaling, but corresponds to a theory with two interacting
scalars in the Einstein frame.
Altogether, this may indicate that the origin
of cosmic acceleration could not purely be the effect of one rolling scalar field, but also 
involves modified gravity effects similar to $f(R)$-theories when quantum fluctuations
of the scalar are taken into account.

We conclude that quantum fluctuations do play an important role 
(i) in coupled dynamical dark energy scenarios
even if we allow fine-tuning in the form of renormalization conditions,
(ii) for the shape of the quintessence potential
and (iii) for its interplay with gravity.

\vspace*{1cm}

\begin{appendix}

%%%%%%%%%%%%%%%%%%%%%%%%%%%%%%%%%%%%%%%%%%%%%%%%%%%%%%%%%%%%%%%%%%%%%%%%%%%%%%%%%%%%%%%%%%%%%%%%%%%%%%%%%%%%%%%%%%%%%%%%%%%%%%%%%%%%%%

\section{\label{app:Ren}Renormalization in Bubble-Approximation}

%%%%%%%%%%%%%%%%%%%%%%%%%%%%%%%%%%%%%%%%%%%%%%%%%%%%%%%%%%%%%%%%%%%%%%%%%%%%%%%%%%%%%%%%%%%%%%%%%%%%%%%%%%%%%%%%%%%%%%%%%%%%%%%%%%%%%%

In section \ref{sec:3} the effective scalar potential has been calculated using a
``multi-bubble" approximation where only diagrams with one vertex but with arbitrary number of loops have been kept.
Here we will show that this effective potential can be consistently renormalized under the following assumptions:
(i) The higher derivatives of the potential $V(\phi)$ can be roughly estimated by $V^{(k+2)}(\phi)=\mathcal{O}(V''(\phi)/M(\phi)^k)$
with the scale height $M(\phi)\equiv (d\ln V''/d\phi)^{-1}$.
(ii) The renormalized value of the one-loop tadpole integral $I_1$, see eq. (\ref{Ik}), can be characterized by a scale $\mu$, i. e.
$I_1^{\rm finite}=\mathcal{O}(\mu^2)$, with $\mu^2\gg V''(\phi)$.
(iii) The renormalization is carried out only up to contributions proportional to $V''(\phi)/\mu^2\ll 1$.

The third assumption is necessary since the corrections to the bubble 
approximation from graphs with at least two vertices are generically also of the order $V''(\phi)/\mu^2$,
and thus would have to be included if terms of this order were to be considered.

The counterterms will in general contain the divergent parts of the one-loop integrals $I_k$.
The splitting 
\begin{equation}
I_k=I_k^{\rm finite}+I_k^{\rm div} \ {\rm for} \  k=0,1,2
\end{equation}
into finite and
divergent parts will depend on the regularization which is used and a specific renormalization scheme.
Here, we will not pick up a special prescription, but perform the renormalization 
in full generality except the assumption that  both the finite and divergent
parts separately obey the relation $I_k(m^2)=(-1)^{k-1}/(k-1)!(d/dm^2)^kI_0(m^2)$, see eq. (\ref{Ik}),
where it is understood that $I_k^{\rm div}\equiv 0$ for $k\geq 3$. This implies
that $I_k^{\rm div}(m^2)$ for $k=0,1,2$ are polynomials in $m^2\equiv V''$ of order $k$,
exactly as required (see also sec. \ref{sec:2}). 
In general, the divergent parts will depend on some regularization parameter
and go to infinity if the regularization is removed. For example, in the case of a cutoff $\Lambda$
one has $I_0^{\rm div}\sim\Lambda^4$, $I_1^{\rm div}\sim\Lambda^2$ and $I_2^{\rm div}\sim\ln\Lambda^2$.
The corrections to the multi-bubble approximation then contain terms of the order $V''/I_1^{\rm finite}\sim V''/\mu^2$
and $V''/I_1^{\rm div}\sim V''/\Lambda^2$, where the latter contribution is suppressed if the cutoff is sent to
infinity. 
Thus, $V''/\mu^2$ is an {\it upper bound} to the corrections 
which we will therefore use throughout the calculations for simplicity,
formally corresponding to the power-counting rule\footnote{This rule can of course only be used to 
determine whether corrections suppressed
by some power of $I_1$ are sub-leading or not,
but not in the actual leading-order calculation.} $I_1\sim\mathcal{O}(\mu^2)$.

For the renormalization procedure we start with a canonically normalized scalar field
with bare potential $V_B(\phi)\equiv V(\phi)+\delta V(\phi)$, which is split up
into the renormalized potential $V(\phi)$ and the counterterms $\delta V(\phi)$.
Since there appears no anomalous dimension in the multi-bubble approximation,
we do not introduce a corresponding (unnecessary) counterterm for simplicity.
Furthermore, we split the counterterms into a series
\begin{equation}\label{CounterLoopSum}
\delta V(\phi) = \sum_{L=1}^\infty \delta V_L(\phi)\hbar^L
\end{equation}
in powers of an order parameter $\hbar$, which will in the end be set to one,
together with the replacement $d^4q\rightarrow\hbar\, d^4q$ in the loop integrals.
Thus we obtain from eq. (\ref{TadSum})
\begin{equation}\label{TadSumWithCounter}
V_{\textit{eff}}''(\phi) = \!
\sum_{N=0}^\infty\!\frac{(V\!+\delta V)^{(2N+2)}}{2^NN!}
\left(\!\int\!\!\!\frac{d^4q}{(2\pi)^4}\frac{\hbar}{q^2+\!V''\!+\delta V''}\!\right)^N \!\!\!.
\end{equation}
The main task is to expand this expression in $\hbar$ using eq. (\ref{CounterLoopSum}).
Diagrammatically, with lines corresponding to an (euclidean) propagator $1/(q^2+V''(\phi))$,
$V_{\textit{eff}}''$ is equal to the bubble
sum as in eq. (\ref{TadSum}) where also diagrams with counterterm-vertices
are added and all propagators can carry an arbitrary number of $\delta V''$-insertions,
denoted by crosses with two or more legs respectively.

The renormalization can now be carried out order by order in $\hbar$.
The contribution of order $\hbar^1$ (``one-loop") is
\begin{equation}\label{VeffWithCounter1L}
V_{\textit{eff}}''(\phi)_{1} = 
\rule{0mm}{7mm}\parbox{11mm}{\includegraphics{tadpole.eps}} 
+ \parbox{11mm}{\includegraphics{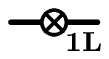}}
\, = \frac{V^{(4)}}{2}I_1(V'')+\delta V_1''.
\end{equation} 
This fully determines the one-loop counterterm
\begin{equation}\label{Counter1L}
\delta V_1''(\phi) = -\frac{V^{(4)}(\phi)}{2}I_1^{\rm div}(V''(\phi)),
\end{equation}
chosen such that
\begin{equation}
V_{\textit{eff}}''(\phi)_{1} = \frac{V^{(4)}(\phi)}{2}I_1^{\rm finite}(V''(\phi)).
\end{equation}
At two-loop order, i.e. at order $\hbar^2$, one gets
\begin{equation}\label{VeffWithCounter2L}
\begin{array}{lcl}
\displaystyle V_{\textit{eff}}''(\phi)_{2} &\displaystyle =& 
\parbox{11mm}{\includegraphics{doubletadpole.eps}} 
+ \parbox{11mm}{\includegraphics{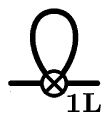}}
+ \parbox{11mm}{\includegraphics{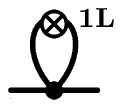}}
+ \parbox{11mm}{\includegraphics{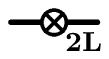}} \\
&\displaystyle =&\displaystyle \frac{V^{(6)}}{8}I_1^2+\frac{\delta V_1^{(4)}}{2}I_1
-\frac{V^{(4)}}{2}\delta V_1''I_2+\delta V_2'',
\end{array}
\end{equation}
where $I_k\equiv I_k(V'')$. 
The counterterm $\delta V_1^{(4)} = \frac{d^2}{d\phi^2}\delta V_1''$
can be calculated from eq. (\ref{Counter1L}),
\begin{equation}\label{Counter1Lderiv}
\begin{array}{lcl}
\displaystyle \delta V_1^{(4)} \!\!\!
&\displaystyle =&\displaystyle -\frac{V^{(6)}}{2}I_1^{\rm div}+V^{(5)}I_2^{\rm div}V'''+\frac{(V^{(4)})^2}{2}I_2^{\rm div} \\
&\displaystyle =&\displaystyle -\frac{V^{(6)}}{2}I_1^{\rm div}
\Bigg[1-\Bigg(2\underbrace{\frac{V^{(5)}V'''}{V^{(6)}}}_{\mathcal{O}(V'')}
+\underbrace{\frac{(V^{(4)})^2}{V^{(6)}}}_{\mathcal{O}(V'')}\!\Bigg)\!\!\!
\underbrace{\frac{I_2^{\rm div}}{I_1^{\rm div}}}_{\sim\mathcal{O}(\frac{I_2^{\rm div}}{\mu^2}\!)}\!\!\!\Bigg] \\
&\displaystyle =&\displaystyle -\frac{V^{(6)}}{2}I_1^{\rm div}\left[1+\mathcal{O}\left(\frac{V''}{\mu^2}I_2^{\rm div}\right)\right],
\end{array}
\end{equation}
where we have used the assumptions and power-counting rules discussed above.
The main contribution comes from the part where $d^2/d\phi^2$ only acts on the coupling $V^{(4)}$
in eq. (\ref{Counter1L}),
whereas the other terms are suppressed by a relative factor of the order $V''/\mu^2\ll 1$.
The logarithmic divergence $I_2^{\rm div}$ belongs to
the sub-leading order and can only be consistently renormalized together with two-vertex graphs
as discussed above. 
For brevity, we will denote any such corrections by 
$\epsilon\equiv V''p(I_2^{\rm div})/\mu^2$ where $p$ stands for a polynomial
with maximally order one coefficients.

The third diagram in eq. (\ref{VeffWithCounter2L}) does not contribute at all in leading order,
as can be seen by comparison with the first one, 
\begin{equation}\label{VeffWithCounterInsertion2L}
-\frac{V^{(4)}}{2}\delta V_1''I_2 = \frac{V^{(6)}}{4}I_1^2\left(\frac{(V^{(4)})^2}{V^{(6)}}\frac{I_2I_1^{\rm div}}{I_1^2}\right)
= \frac{V^{(6)}}{4}I_1^2\,\mathcal{O}\left(\epsilon\right),
\end{equation}
where we have again used the power-counting rules.
Thus, using eqs. (\ref{VeffWithCounter2L},\ref{Counter1Lderiv},\ref{VeffWithCounterInsertion2L}) gives
\begin{equation}
V_{\textit{eff}}''(\phi)_{2} = \frac{V^{(6)}}{8}\left[I_1^2-2I_1I_1^{\rm div}\right]\left(1+\mathcal{O}(\epsilon)\right) + \delta V_2''.
\end{equation}
The ``nested" divergence proportional to $I_1^{\rm finite}I_1^{\rm div}$ thus cancels as required and
the two-loop counterterm and renormalized effective potential can be determined. They are given by
the special case $L=2$ of the general ansatz
\begin{eqnarray}
\delta V_L''(\phi) &=& (-1)^L\frac{V^{(2L+2)}}{2^LL!}\left(I_1^{\rm div}(V'')\right)^L\left(1+\mathcal{O}(\epsilon)\right)
 \ \qquad \label{CounterL}\\
V_{\textit{eff}}''(\phi)_{L}\! &=& \frac{V^{(2L+2)}}{2^LL!}\left(I_1^{\rm finite}(V'')\right)^L\left(1+\mathcal{O}(\epsilon)\right).
\label{VeffRenormL}
\end{eqnarray}
For $L\geq 3$, the upper eqs. can be proven by induction. Let us assume $\delta V_l''(\phi)$ is given by eq. (\ref{CounterL})
for $l\leq L-1$. Then we have to show eqs. (\ref{CounterL},\ref{VeffRenormL}) for $L$, which is done
by evaluating the contribution of order $\hbar^L$ to $V_{\textit{eff}}''$ as given in eq. (\ref{TadSumWithCounter}).
This contribution can also be expressed as a sum of diagrams similarly to eqs. (\ref{VeffWithCounter1L},\ref{VeffWithCounter2L}).
Let $\mathcal{G}(N,l,n_1,n_2,\dots,n_{L-1})$ denote the sum of all diagrams with $N$ tadpoles, $n_k$ insertions
of $\delta V''_k$ into internal lines and with vertex $V^{(2N+2)}$ if $l=0$ and $\delta V_l^{(2N+2)}$ if $l\geq 1$.
Since the diagram should be of order $\hbar^L$, one has $L=N+l+\sum kn_k$. This immediately implies
that only counterterms of order less than $L$ can enter for $N\geq 1$. All these contributions
can be calculated using the ansatz (\ref{CounterL}). For this we note that the higher derivatives
of $\delta V_k''$ from eq. (\ref{CounterL}) have the form
\begin{equation}
\delta V_k^{(m)} = (-1)^k\frac{V^{(2k+m)}}{2^kk!}\left(I_1^{\rm div}(V'')\right)^k\left(1+\mathcal{O}(\epsilon)\right)
\end{equation}
in leading order in $\epsilon$, which can be seen in a similar way as in eq. (\ref{Counter1Lderiv}).
Furthermore, using this relation, one can see that among the diagrams with fixed $N$ and $L$
all diagrams with counterterm insertions are suppressed by at least one factor of order $\epsilon$
relative to $\mathcal{G}_N\equiv\mathcal{G}(N,L-N,0,\dots,0)$, as in the two-loop case, 
see eqs. (\ref{VeffWithCounter2L},\ref{VeffWithCounterInsertion2L}).
Thus we obtain in leading order 
\begin{equation}
\begin{array}{l}
\displaystyle {V_{\textit{eff}}''}_{\, L}  = 
\mathcal{G}_0 + \sum_{N=1}^{L-1}\mathcal{G}_N\left(1+\mathcal{O}(\epsilon)\right) + \mathcal{G}_L \\
\displaystyle \quad =
\rule{0mm}{10mm}\parbox{11mm}{\includegraphics{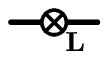}} 
+ \parbox{11mm}{\includegraphics{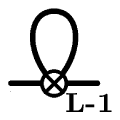}} 
+ \parbox{11mm}{\includegraphics{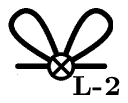}} 
+ \cdots + \parbox{11mm}{\includegraphics{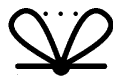}} \\
\displaystyle\quad =
\delta V_L'' + \sum_{N=1}^{L-1}\frac{\delta V_{L-N}^{(2N+2)}}{2^NN!}I_1^N(1+\mathcal{O}(\epsilon)) + \frac{V^{(2L+2)}}{2^LL!}I_1^L\\
\displaystyle\quad =
\delta V_L'' + \frac{V^{(2L+2)}}{2^LL!}\sum_{N=1}^{L}\!{L\choose N}\!I_1^N(-I_1^{\rm div})^{L-N}(1+\mathcal{O}(\epsilon)) \\
\displaystyle\quad =
\delta V_L'' + \frac{V^{(2L+2)}}{2^LL!}\left((I_1^{\rm finite})^L-(-I_1^{\rm div})^L\right)(1+\mathcal{O}(\epsilon)).
\end{array}
\end{equation}
The last line shows that again all mixed terms in $I_1^{\rm finite}$ and $I_1^{\rm div}$ cancel,
which is important for consistency, and implies the validity of the ansatz in eqs. (\ref{CounterL},\ref{VeffRenormL})
at order $L$, thereby completing the proof. Thus the final result for the renormalized second derivative of the
effective potential
in bubble-approximation is  given by the sum over all contributions from eq. (\ref{VeffRenormL}) ($\hbar\equiv 1$),
\begin{equation}
V_{\textit{eff}}''(\phi) = \sum_{L=0}^\infty \frac{V^{(2L+2)}(\phi)}{2^LL!}\left(I_1^{\rm finite}(V''(\phi))\right)^L(1+\mathcal{O}(\epsilon)).
\end{equation}

%%%%%%%%%%%%%%%%%%%%%%%%%%%%%%%%%%%%%%%%%%%%%%%%%%%%%%%%%%%%%%%%%%%%%%%%%%%%%%%%%%%%%%%%%%%%%%%%%%%%%%%%%%%%%%%%%%%%%%%%%%%%%%%%%%%%%%

\section{\label{app:HKE}Heat Kernel Expansion}

%%%%%%%%%%%%%%%%%%%%%%%%%%%%%%%%%%%%%%%%%%%%%%%%%%%%%%%%%%%%%%%%%%%%%%%%%%%%%%%%%%%%%%%%%%%%%%%%%%%%%%%%%%%%%%%%%%%%%%%%%%%%%%%%%%%%%%

The one-loop contribution to the effective action for the Lagrangian (\ref{LagrangeCurved}) is
given by the determinant
\begin{equation}
\Gamma[\phi]_{1L} = -\frac{1}{2i}\ln\det\hat A,
\end{equation}
with the operator $\hat A\equiv \Box + X$
and $X=\partial^2V(\phi,R)/\partial\phi^2$.
The generalized zeta-function for $\hat A$ is $\zeta_A(\nu)\equiv\sum_m\lambda_m^{-\nu}$
where ${\lambda_m}$ are the eigenvalues of $\hat A$.
Using zeta-function regularization (see e.g. \cite{Birrell:1982ix,Hawking:1976ja}) the determinant
can be written as 
\begin{equation}\label{EffActZetaReg}
\Gamma[\phi]_{1L} = \frac{1}{2i}(\zeta_A'(0)+\zeta_A(0)\ln\mu^2),
\end{equation}
where we introduced a renormalization scale $\mu$.
The zeta-function can also be expressed via the heat kernel $K(x,y,s)$
fulfilling the heat equation $i\frac{\partial}{\partial s}K(x,y,s)=\hat A(x)K(x,y,s)$
with $K(x,y,0)=\delta(x-y)$ as
\begin{equation}\label{ZetaAndHK}
\zeta_A(\nu) = \frac{i}{\Gamma(\nu)}\int_0^\infty d\!s (is)^{\nu-1} \int d^4\!x K(x,x,s). 
\end{equation}
The ansatz for $K$ of Refs. \cite{Jack:1985mw,Parker:1984dj} is
\begin{equation}
\begin{array}{lcl}
\displaystyle K(x,y,s) &\displaystyle =&\displaystyle\frac{i}{(4\pi is)^2}\, \Delta_{VM}^{1/2}(x,y)\, \bar G(x,y,s) \\
&&\displaystyle \exp\left(-\frac{\sigma(x,y)}{2is}-is\left(X(y)-\frac{R(y)}{6}\right)\right),
\end{array}
\end{equation}
where $\sigma(x,y)$ is the proper arclength along the geodesic from $x$ to $y$
and $\Delta_{VM}$ the Van Vleck-Morette determinant fulfilling $\Delta_{VM}(x,x)=-g(x)$.
Inserting this ansatz together with the expansion $\bar G(x,y,s)=\sum_{j=0}^\infty(is)^j\bar g_j(x,y)$
into eq. (\ref{ZetaAndHK}) and using eq. (\ref{EffActZetaReg}) yields for the effective action
\begin{equation}\label{EffActCurved}
\begin{array}{lcl}
\displaystyle \Gamma[\phi]_{1L} &\displaystyle = &\displaystyle\int\!\!\frac{d^4\!x}{32\pi^2}\sqrt{-g}
\left[-\frac{\tilde X^2}{2}\left(\ln\frac{\tilde X}{\mu^2}-\frac{3}{2}\right) \right.\\
&&\displaystyle \left. -\bar g_2(x,x)\ln\frac{\tilde X}{\mu^2} 
 +\sum_{j=3}^\infty\bar g_j(x,x)\frac{(j-3)!}{\tilde X^{j-2}}\right],
\end{array}
\end{equation}
where we have set $\tilde X\equiv X-R/6$. The coincidence limits of the $\bar g_j$
can be calculated recursively. We quote the result for the lowest orders from Ref.~\cite{Jack:1985mw},
\begin{equation}\label{gj}
\begin{array}{lcl}
 \bar g_0(x,x) &=& 1\,,\  \bar g_1(x,x) \ =\ 0, \\
 \bar g_2(x,x) &=& \frac{1}{180}(R_{\mu\nu\rho\sigma}R^{\mu\nu\rho\sigma}\!-\!R_{\mu\nu}R^{\mu\nu})
-\frac{1}{30}\Box R+\frac{1}{6}\Box X \\
& =& \frac{1}{120}C-\frac{1}{360}G -\frac{1}{30}\Box R+\frac{1}{6}\Box X,
\end{array}
\end{equation} 
where $C$ and $G$ are the Weyl- and Gauss-Bonnet terms as given in section \ref{sec:4}.
The $\bar g_j$ with $j\geq 3$ contain higher-order curvature scalars built from the curvature- and
Ricci tensors and space-time derivatives of $R$ and $X$ and correspond to finite contributions
to the one-loop effective action (\ref{EffActCurved}), whereas the $j=0,1,2$-contributions
come along with divergences proportional
to $\bar g_0\tilde X^2$, $\bar g_1 \tilde X$ and $\bar g_2$. Using eq. (\ref{gj}) one can see
that it is necessary to introduce counterterms proportional to $\tilde X^2=(\partial^2V/\partial\phi^2-R/6)^2$,
$\Box(X-R/5)=\Box(\partial^2V/\partial\phi^2-R/5)$, $C$ and $G$ in order to cancel these divergences, which is already
done implicitly in the result (\ref{EffActCurved}) for the effective action through
the zeta-function regularization \cite{Hawking:1976ja}. Nevertheless,
all operators contained in the counterterms should be already present
in the tree level action.

\end{appendix}

\begin{acknowledgments}
The author would like to thank Florian Bauer as well as Manfred Lindner for useful comments and discussions. 
This work was supported
by the {}``Sonderforschungsbereich 375 f\"{u}r Astroteilchenphysik
der Deutschen Forschungsgemeinschaft''. 
\bibliographystyle{apsrev}
\bibliography{references-QuQuint}

\end{acknowledgments}

\end{document}